\documentclass[aps,nofootinbib,onecolumn,notitlepage,superscriptaddress,11pt]{revtex4-2}

\usepackage{amssymb, amsmath, amsthm, amsfonts}
\usepackage{graphicx}
\usepackage{color}
\usepackage{mathrsfs}
\usepackage{physics}
\usepackage{mathtools}

\DeclareMathOperator{\E}{\mathbb{E}}

\usepackage[UKenglish]{isodate}
\usepackage[UKenglish]{babel}

\usepackage{hyperref}

\begin{document}

\title{On the Map-Territory Fallacy Fallacy}

\author{Maxwell J D Ramstead}
\email{maxwell.ramstead@verses.io}
\affiliation{VERSES Research Lab, Los Angeles, CA, 90016, USA}
\affiliation{Wellcome Centre for Human Neuroimaging, University College London, London, UK}

\author{Dalton A R Sakthivadivel}
\email{dalton.sakthivadivel@stonybrook.edu}
\affiliation{VERSES Research Lab, Los Angeles, CA, 90016, USA}
\affiliation{Department of Mathematics, Stony Brook University, Stony Brook, NY, USA}
\affiliation{Department of Physics and Astronomy, Stony Brook University, Stony Brook, NY, USA}
\affiliation{Department of Biomedical Engineering, Stony Brook University, Stony Brook, NY, USA}

\author{Karl J Friston}
\email{k.friston@ucl.ac.uk}
\affiliation{VERSES Research Lab, Los Angeles, CA, 90016, USA}
\affiliation{Wellcome Centre for Human Neuroimaging, University College London, London, UK}

%%%%%%%%%%%%%%%%%%%%%%%%%%%%%%%%%%%%%%%%%%%%%%%%%%%%%%%%%%%%%%%%%

\date{\today}

\begin{abstract}

This paper presents a meta-theory of the usage of the free energy principle (FEP) and examines its scope in the modelling of physical systems. We consider the so-called `map-territory fallacy' and the fallacious reification of model properties. By showing that the FEP is a consistent, physics-inspired theory of inferences of inferences, we disprove the assertion that the map-territory fallacy contradicts the principled usage of the FEP. As such, we argue that deploying the map-territory fallacy to criticise the use of the FEP and Bayesian mechanics itself constitutes a fallacy: what we call the \textit{map-territory fallacy fallacy}. In so doing, we emphasise a few key points: the uniqueness of the FEP as a model of particles or agents that model their environments; the restoration of convention to the FEP via its relation to the principle of constrained maximum entropy; the `Jaynes optimality' of the FEP under this relation; and finally, the way that this meta-theoretical approach to the FEP clarifies its utility and scope as a formal modelling tool. Taken together, these features make the FEP, uniquely, \textit{the} ideal model of generic systems in statistical physics.

\end{abstract}

\maketitle

%%%%%%%%%%%%%%%%%%%%%%%%%%%%%%%%%%%%%%%%%%%%%%%%%%%%%%%%%%%%%%%%%

\section{Introduction}

Bayesian mechanics and the free energy principle (FEP) have been proposed as a universal modelling methodology for physical things, and in particular, for self-organising systems \cite{friston2022free, Friston2019, DaCosta2021}. The FEP is a mathematical statement (or indeed a physical principle) which provides a unifying perspective on formal approaches to behaviour and dynamics \cite{Friston2010}. Bayesian mechanics, a corollary of the FEP, comprises a set of tools to write mechanical theories (i.e., laws of motion) that relate the dynamics of states or paths of a physical system to the dynamics of probabilistic beliefs (or conditional probability densities) that are encoded by some (so-called internal) subsets of the states of the system \cite{ramstead2022bayesian}. These internal states (or their dynamics) are coupled to other (external) subsets of the system in such a way that they can be read as parameters for probability densities over the external subset. Bayesian mechanics unifies first-principles-based modelling efforts in physics by integrating the FEP with one of the cornerstones of contemporary physics, namely, the constrained maximum entropy principle (CMEP) \cite{Sakthivadivel2022b}. As a consequence, these are tools that allow us to model any particle, or `thing'\textemdash in the sense that it exists as a separable, identifiable thing \cite{fields2021free, sakthivadivel2022d}\textemdash as if it were inferring the causes of its sensory perturbations and (re)acting so as to remain in system-like configurations. This inference proceeds by the minimisation of a free energy functional of such beliefs as parameterised by internal states, which guarantees\textemdash or describes\textemdash the minimisation of the surprisal of particular states. Using this approach, we can model the dynamics (i.e., the observable behaviour) of things or particles as realising a path of least surprisal. Bayesian mechanics formalises the truism in physics that, on average, systems will do what it is that they do on average. Behind this truism lies a formal apparatus that enables us to write many of the core formulations of contemporary physics in a parsimonious manner. The hope is that this will allow one to take some first steps towards a mechanistic explanation of how self-organisation arises in physical systems.

In this paper, we begin the development of a \textit{bona fide} meta-theory, and philosophy for the usage, of the FEP. A good starting point for this project would be the so-called \textit{map-territory fallacy}, which some have alleged is committed by the architects of the FEP \cite{Bruineberg2020, VanEs2020}. Indeed, some scholars have argued that FEP-theoretic modelling conflates the metaphorical `map' (i.e., the scientific model that scientists use to make sense of some target phenomenon) and `territory' (i.e., the actual natural system that is being modelled). See \cite{andrews2022making} and \cite{ramstead2022bayesian} for a related critical discussion of such claims. On this view, using the technology of FEP (and Bayesian mechanics more broadly) to claim that things or particles engage in inference constitutes a case of model reification. That is, some have alleged that, in describing the dynamics of physical systems as inference, FEP theorists mistakenly conflate their metaphorical `map' of the territory (i.e., the scientific FEP-theoretic or Bayesian mechanical model, which is a set of mathematical structures) with the territory or target system itself, thereby reifying aspects of their model into what are now assumed to be real features of the territory. 

We will argue that deploying the map-territory fallacy to criticise the FEP and Bayesian mechanics in this manner itself constitutes a fallacy: what we will call the \textit{map-territory fallacy fallacy}. We believe that the map-territory fallacy is fallacious for two main reasons. 

Firstly, we will argue that the FEP commits one to an unproblematic\footnote{The story in this paper rests on a deflated and naturalised account of representation and semantics, which is reviewed in \cite{wiese, Ramstead2020semantics, Kiefer2020psycho}. As we detail below, any system that can be modelled as an estimator i.e., as engaging in the statistical estimation of some quantity, is a representation of the estimated system. We are aware that proponents of anti-representationalist philosophy of mind might want to resist this notion. Below, we explain why this resistance is not justified, mathematically speaking.} nested representationalist account of physical dynamics: mathematically, there simply is no conflation of map and territory, i.e., of modelled system and scientific model (see also \cite{Ramstead2019enactive, ramstead2022bayesian, ramstead2021empire, andrews2022making}). Indeed, the technology of the FEP may be used to write down a statistical (generative) model of the states of a system (or paths of a system through its state space), which constitutes \textit{our} scientific models of the target system (as modellers). The model that we construct contains a partition of the states or paths of the system (a `particular' partition, i.e., into particles); this partition captures relations of conditional independence relations amongst subsets of the system, which track each others’ statistical structure dynamically. These subsets of states engage in inference about (or represent the statistical structure of) other subsets of states. Thus, we can think of the FEP, metaphorically, as a map of that part of the territory that \textit{behaves as if it were a map}. Crucially, at this level of analysis, there is no conflation between our FEP-theoretic model and the statistical model that is embodied by a particle in a self-organising system: under the assumption that systems model their environments, we can understand their dynamics in virtue of the model they hold. As such, when we use the FEP to model the way a system models its environment, we simultaneously model the system itself in a high-fidelity manner. This duality is key to dismissing the map-territory fallacy and is an \emph{essential} feature of the success of the FEP. That this exists in the FEP by consequence of the fundamental posits of the theory, in particular, a symmetric boundary across which bidirectional interactions occur, is no accident. 

Secondly, we will argue that, when deployed in the context of FEP-theoretic modelling, the map-territory fallacy rests on some rather deep misunderstandings about the nature of modelling in physics, and about the existence and nature of constraints in modelling physical systems in general. Specifically, we leverage the mathematical fact that, via the principle of maximum entropy, an FEP-theoretic model of belief updating is \textit{the optimal model of systemic dynamics}, to argue that the FEP provides ultimate constraints on what it means to be an optimal (physical, i.e., embodied) model of some target physical phenomenon. 

This allows for the following: we can leverage this fact to claim that the `thing' we are modelling can also be expected to use the principle of maximum entropy to model its environment; since the FEP can be read as the application of the CMEP to model maximum entropy agents modelling their environment, then, referencing the first point, we can produce the high-fidelity model that we have indicated. All this means is that, under the assumption that systems model their environments optimally given some constraints on that model, they will model their environment using the FEP or CMEP. In turn, this entails that we can understand their dynamics in virtue of the model they hold by ourselves using the FEP to reproduce that model. More particularly, we will always reproduce a map of that part of the territory that makes maps.

Additionally, we argue that (in a metaphorical sense) the FEP and Bayesian mechanics also provide a map of any possible map whatsoever of, or held by, a physical system. This undermines any strict notion that maps and territories must always be held distinct conceptually; and thus defeats the map-territory fallacy (at least as applied to the FEP). However, this is not to suggest that the map and the territory are never conflated; they obviously are in cases of genuine model reification \cite{andrews2022making}. Instead, we suggest that this leads, in a Kantian manner (in terms of logically necessary preconditions for sense-making, and thus the parsing of communicated sensory streams on which cognition is contingent \cite{Ramstead2020semantics}), to constraints on possible maps, which arise from what it means to be a map or model of a physical process at all. 

We first present a technical introduction to the FEP and Bayesian mechanics, and examine the claim that the FEP is, metaphorically, a map of that part of the territory that behaves as if it were a map. It turns out that most things behave, at some level of description, like maps in this sense. We defuse the map-territory fallacy at a first level of analysis, showing that there is no simply ambiguity between model and modelled system in the core FEP formalism. We then discuss the duality of the FEP and the CMEP, and examine the implications of this duality for modelling physical systems. We further defuse the map-territory fallacy regarding the FEP by noting that the FEP/CMEP duality entails constraints on what counts as an optimal model of a physical process. Here, the very distinction between map and territory is blurred; and the metaphor itself is thereby shown to be less useful than might have been hoped initially. Finally, we summarise our critical discussion and open onto future research questions that ought to be taken up by philosophers of the FEP. 

Altogether, we present a thesis that modelling internal states modelling external states constitutes the best possible model of the internal states of a `thing,' by proving that
\begin{enumerate}
    \item things with internal states model external states, and that this model-of-models is captured by the FEP (Sections \ref{intro-to-fep} and \ref{map-behaves-map}), 
    \item the FEP is an optimal model at the particle-level (Section \ref{optimal-section-i}),
    \item we can achieve a relevant model of those internal states by leveraging particle-environment symmetry (Section \ref{dual-section}), and
    \item abiding by this symmetry, the FEP-theoretic model of the beliefs of a particle is the optimal model at the modeller-level (Section \ref{optimal-section-ii}).
\end{enumerate}
Together these justify the \emph{utilisation} of the FEP at a semi-axiomatic level, and lay out a set of schema driving its applicability.

\section{The FEP and Bayesian mechanics}

\subsection{Introduction to the FEP}\label{intro-to-fep}

The FEP can be summarised as a principle of least surprisal \cite{friston2022free, ramstead2022bayesian, classical-physics} for systems that possess a particular partition. A particular partition simply takes a system and separates it into a set of internal states and external states by an intervening set of blanket states (also called a \emph{Markov blanket}). This separation will be central to what follows in two senses. Firstly, it is an existential imperative for anything. In other words, if the states of something could not be individuated from everything else, the thing would not exist. Secondly, when talking about maps\textemdash in the formal sense of a function or morphism\textemdash we require the domain and codomain (i.e., the image) of the map to exist. Therefore, the use of a map in this setting presupposes the existence of a partition.

Formulated most generally, the FEP says that the paths taken by a system through its state or phase space are characterised by a Lyapunov function, a function that varies systematically with its dynamics. The utility of Lyapunov functions is they allow us to understand those dynamics varying with such a function as a gradient flow on the function\textemdash in other words, it is crucial to understanding the `inference' component of surprisal minimisation. Under the FEP, this Lyapunov function is surprisal or variational free energy, depending on the setting. Surprisal is simply negative log probability. Variational free energy is a tractable (i.e., easily computed) upper bound on surprisal. Minimising surprisal implies that the variational free energy is minimised, and likewise, minimising the variational free energy places an upper bound on the surprisal. Though not a minimum \emph{per se}, minimising the variational free energy is a necessary first step, and a good proxy, for surprisal minimisation.\footnote{In the limit of exact Bayesian inference, when the functional form of posterior beliefs encoded by internal states coincides with the posterior density over external states, the variational free energy is just the surprisal. See e.g. \cite[Lemma 4.2]{Sakthivadivel2022b}.} That the dynamics of a system encode an inference whenever they minimise this quantity is the consequence of reading surprisal as a Lyapunov function for those dynamics. This gradient flow is what we identify as actually being the process (i.e., the performance) of inference.

As stated, the FEP partitions a `system' into three sets of states: the particular states of a `thing' or `particle' (which is equipped with internal states $\mu$), an environment (which possesses external states $\eta$, relative to the thing in question), and an interface separating but coupling the two (the so-called Markov blanket, $b$), which mediates the influence of $\eta$ on $\mu$ and \emph{vice-versa}. When we say `recognition model' or `Bayesian belief' we mean a parameterised probability density over external states conditioned on particular states $\mu$ and $b$. The \emph{variational} free energy is based on a relative entropy, measuring the divergence between a particle's recognition model of its environment and the true posterior conditional density over external states given blanket states. The recognition model is parameterised by (the dynamics of) internal states in such a way that the particle synchronises with the environment. More specifically, suppose there exists a parametric model of the probability of external states, $q(\eta; \hat\eta_b, \Sigma_{\eta\mid b})$, where we have parameterised the recognition model by the average\footnote{In fact this is a conditional average, since the state of the environment is dependent on the state of the boundary connected to it; hence we denote this as $\hat\eta_b$. More generally, the parameter controlling the peak of the density\textemdash the conditional \emph{maximum a posteriori} estimate\textemdash will do.} state $\hat\eta_b$ and variance $\Sigma_{\eta\mid b}$ of the environment. 

If there exists a boundary of the form of a Markov blanket, then there exists a function $\sigma$ relating (the dynamics of) internal states to the environment across the boundary \cite[Lemma 4.3]{Sakthivadivel2022b}. In that case, we can convert these parameters to $\sigma(\hat\mu_b)$ and $[\eta - \sigma(\hat\mu_b)]^2$, meaning that the \emph{particle} models its environment. Ultimately this simply means that, in virtue of being coupled (vicariously, i.e., via blanket states) to its environment, the particle reflects data about its environment\textemdash or, what it believes that its environment is like. Under the assumption that a particle which exists in an environment models that environment, free energy is implicitly a measurement of coherence. If the particle stops modelling its environment in a particular way (i.e., for a particular $\sigma$), it must have ceased to exist with the mean that $\sigma$ was constructed for; conversely, if a particle ceases to exist, it will eventually stop modelling its environment. The variational free energy measures how far this recognition model of what lies beyond the blanket is from the true probability density over external states, given blanket states. If the recognition model is bad, and the free energy is high, it is likely that the system will move into surprising (i.e., implausible or uncharacteristic) states; this would render it non-existent, in the sense that it would be occupying states that were not characteristic of the thing in question (e.g., a fish out of water). These statements tie together the reasoning that things which exist in an environment reflect the statistics of their environment in a particular way based on the way they are coupled to their environment, or, they do not exist that way.

Mathematically, we can express this by observing the following: since the joint probability $p(\eta, b, \mu)$ factorises into $p(\eta \mid b, \mu)p(b, \mu)$ (see the probabilistic chain rule) and logarithms decompose additively, we have 
\begin{align}\label{vfe-eq}
\int &q(\eta; \sigma(\hat\mu_b), \Sigma_{\eta \mid b})\log q(\eta; \sigma(\hat\mu_b), \Sigma_{\eta \mid b}) \dd{\eta} -\int q(\eta; \sigma(\hat\mu_b), \Sigma_{\eta \mid b}) \log p(\eta, b, \mu) \dd{\eta} = \nonumber \\ & \int q(\eta; \sigma(\hat\mu_b), \Sigma_{\eta \mid b}) \log q(\eta; \sigma(\hat\mu_b), \Sigma_{\eta \mid b}) \dd{\eta} - \int q(\eta; \sigma(\hat\mu_b), \Sigma_{\eta \mid b}) \log p(\eta \mid b, \mu) \dd{\eta} - \log p(\mu, b)
\end{align}
where the left-hand side measures a divergence between $q(\eta; \sigma(\hat\mu_b), \Sigma_{\eta \mid b})$ and the true joint density; whilst on the right-hand side, the divergence between that recognition model and the true \emph{conditional} density on external states bounds the surprisal of internal states from above. (Note that this difference term, which we will denote $F$, is always non-negative, since it is a KL divergence\textemdash as such, it is \emph{added} to the surprisal in general. From that we deduce $- \log p(\mu, b) + F \geq - \log p(\mu, b)$.) 

The FEP can be used to reformulate the least action principles that appear in both classical and quantum physics, due to intimate connections between physical law and probability \cite{Friston2019, fields2021free}. In contemporary physics, classical laws are actually the deterministic limit of underlying probabilistic specifications of the behaviours of a system; indeed, we can quantify the degree of divergence from the most probable path of a system by keeping track of quantum effects. The classical path of least action is, definitionally, the most probable path a system will take, i.e., in the Gaussian case, the average path, and thus classical physics is quantum physics modulo what one might call randomness.\footnote{Whilst different approaches to quantum mechanics emphasise different aspects of the theory, this is especially evident in the path integral formulation of quantum physics, which is reflected in the path integral formulation of Bayesian mechanics, cf. \cite{classical-physics}.}

We have known since Prigogine that we can always reformulate the physical laws that pertain to systemic dynamics in terms of surprisal (and related metrics, such as entropy) \cite{Prigogine1978, crooks}. That is, we can always take laws of physics formulated in terms of probability densities, as just described, and reformulate them in terms of surprisal and entropy. In this setting, rather than speak of the most probable path or state, one speaks of the least surprising one, where surprisal is parameterised by some `attractor state.' This is not specific to FEP-theoretic modelling; it is a hallmark of all contemporary approaches to physics.

Bayesian mechanics explains why every physical thing looks as if it were performing inference over the causes of its perturbations, forming (Bayes) optimal beliefs about the causes of those perturbations \cite{Sakthivadivel2022b}. The interesting thing about this is that surprisal is suited to act as a kind of `ontological potential' for the system, in the sense that the surprisal is equivalent to a constraint on what states are accessible to the system; the trajectories of a system are curves on this surprisal, which satisfy some equation of motion derived from the surprisal itself. As has been asserted in recent literature \cite{classical-physics}, one reading of the FEP would hold that the crux of the ubiquity of the FEP is that every\emph{thing} which is coupled to another thing holds a model of that thing in the following very trivial sense: in virtue of this coupling, one system estimates the parameters of the probability density over states of the other system (and vice-versa). Since the FEP regards these systems as estimators, they are performing inference in the sense of parametric Bayesian inference. Given that the term `belief' is quite theoretically loaded (especially in philosophy), we ought perhaps to say `estimator,' instead. Once more, our contention is that anything which estimates the parameters of a density is engaged in a form of Bayesian inference, by mathematical definition; i.e., it is mathematically indisputable that systems that estimate statistics represent, or encode, or otherwise hold probabilistic beliefs\textemdash whether the practitioner likes the word `beliefs' or not, and irrespective of what commitments one makes with encodings and representations.  

There are two main ways to implement FEP-theoretic technology to model the dynamics of a system; with the former being the asymptotic limit of the latter (for detailed discussion, see \cite{ramstead2022bayesian, classical-physics}). First, we can formulate the probability density dynamics of a system \cite{Friston2019, Sakthivadivel2022b}. In this setting, surprisal quantifies how unlikely it is that the system will visit some states, given the kind of thing that it is. The least surprising states are those that are characteristic of the particle. Second, we can formulate the FEP in terms of paths \cite{friston2022free, classical-physics}. In this formulation of the FEP, the surprisal quantifies how far a path deviates from the expected path. For instance, under constraints that reproduce the correspondence principle of quantum physics, the least surprising path is simply the classical path; this is like saying that particles in quantum physics do inference over the forces acting on them at a macroscopic level and act accordingly on average. That this higher-level inference is generally conducive to self-organisation has been remarked on in \cite{neural-reps}.

\subsection{A map of that part of the territory that behaves as if it were a map}\label{map-behaves-map}

We will now argue that the FEP evinces a `map' of sorts: a map of that part of the territory which behaves as if it were a map. In other words, the FEP provides us with tools to understand mathematically the representational capacities of self-organising systems. Even more simply, it gives us tools to model systems that look as if they are modelling the world, producing a map of the mapmaker, as it were. 

How does one get beliefs and inference from FEP-theoretic models? How does FEP-theoretic technology allow us to model self-organising systems as themselves engaging in inference and belief updating? The joint probability $p(\eta, b, \mu)$ plays a central role here. This generative model can be read as a probabilistic specification of the states characteristically occupied by the joint particle-environment system. In this sense, it provides constraints on the kind of states the joint system can be found in. In other words, it is the `territory' that characterises the system. On another reading, it can be regarded as a generative model that gives rise to the free energy in \eqref{vfe-eq}, and thereby the free energy gradients that underwrite particular dynamics. On this view, it is a generative model that is entailed by the recognition model, where we can read the recognition model as a map; namely, a map from internal states to Bayesian beliefs about external states (i.e., the probabilistic image of the mapping). If we now move to a meta-theoretical perspective and consider one particle (e.g., a scientist or philosopher) observing another particle, we have the interesting situation where the philosopher may impute a particular generative model that best explains the mechanics of the observed particle. Is this meta-model a map or a territory?\footnote{This meta-theoretical move is commonplace in practical applications of the FEP and is sometimes referred to as meta-Bayesian inference or, more simply, observing the observer \cite{daunizeau}. Practically, it involves optimising the parameters of a generative model, such that under the FEP (the ideal Bayesian observer assumption) the observable behaviour of the particle is rendered the most likely. This application of the FEP is often described as computational phenotyping \cite{schwartenbeck}.}

The meta-model is our scientific model of the system, as observers or modellers. We generally assume that the system is its `own best model.' In other words, our scientific model is just the joint probability density that describes the state of the system (i.e., the territory). This can be read as a new take on the \emph{nouvelle AI} idea that physical systems are their own best models; or as a new take on the good regulator theorem. For instance, in work on robotics following the tradition of embodied cognition \cite{beer1995computational, brooks1991intelligence}, practitioners have eschewed the construction of agents with internal representations of some environment. In these systems, the physics and geometry of the situation were sufficient to endow these agents with the capacity to couple to an environment. The world is, on this view, its own best representation. In our view, this approach dovetails nicely with FEP-theoretic modelling, where the generative model in FEP-theoretic constructions is just a joint probability density defined over all the states (or paths) of a system. In other words, the generative model is just our representation of the world as we believe it to be; and this model need not be encoded directly in the particle or agent. Given a generative model or Lagrangian of the appropriate sort, we can show that the system evinces a particular partition (i.e., contains particles); and we can show that subsets of the system track each other, where tracking means inferring or becoming the sufficient statistics of probabilistic beliefs about their external states. We interpret this tracking as a form of inference, namely, variational or approximate Bayesian inference under a generative model (i.e., the map).

It is important to say explicitly that the sense in which a system is its own best model (\emph{qua} generative model) is not the same sense in which particles of the system are models (i.e., maps) of their environment, or of themselves acting in their environments (\emph{qua} recognition model); they are of course connected, in the sense that the dynamics of internal states or paths (and thereby, of particular recognition models) entail a generative model (via \eqref{vfe-eq}, for instance)  \cite{Ramstead2019enactive}. 

It is important to defuse a possible (but in our view na\"ive) objection. It is a truism that physical systems need not explicitly calculate their trajectories of motion, to be modelled as pursuing such trajectories. In developing a Bayesian mechanical account of the dynamics of systems, we need not assume that the particle itself is literally performing inference.\footnote{Ways of defending stronger, increasingly literal versions of this positions are available as well. For a defence of the claim that physical systems are quite literally in the business of inference, see \cite{Kiefer2017literal, Kiefer2020psycho}. Related to this, it is possible to articulate a version of the notion of notion of implementation, whereby a physical process implements a computation provided that a so-called interpretation map exists, and commutes in the appropriate manner, such that the physical process can be interpreted as implementing a computation; see \cite{horsman2014does} for the general account, and \cite{fields2021free, fields2022neurons} for its application to the FEP.} What is at stake is an `as if' description. We simply assume that there exists a quantity that varies systematically with the dynamics of the system. It just so happens that, under the FEP, this quantity turns out to be \textit{surprisal} (or, equivalently, variational free energy); and that the minimisation of this quantity is mathematically equivalent to inference, in the sense of being an estimator, as discussed above. 

Thus, \textit{our} map (as scientists and modellers) can be identified in an unproblematic way as the generative model or Lagrangian of the system, which FEP-theoretic technology enables us to write down. And it so happens that our model is a model of the \textit{representational capacities of subsets of the target system}: that is, our map is precisely a map of systems that behave as if they were maps, allowing us to construct scientific models of the maps that are encoded by the internal subset of the system considered. 

In summary, our scientific model is, metaphorically, a map that allows us to say that some (internal) subset of the system looks as if it possesses a map; which we interpret formally as tracking the statistics of another (external) subset. This arguably defuses the first aspect of the map-territory fallacy: namely, it shows that there is no conflation of the map and the territory, at a first level of analysis. 

\subsection{Maps that are optimal employ the CMEP; maps of maps that are optimal employ the FEP}\label{optimal-section-i}

Mathematically, entropy is the unique functional of probability densities which yields consistent, unbiased inferences \cite{SJ, dill1} and is sufficiently general to recover all of equilibrium thermodynamics \cite{jaynes1, dill1, dill2} and much of information and probability theory \cite{jaynes2, jizba2020shannon}. Maximum entropy inference, in which we find the probabilistic model that maximises entropy, given some specific constraints (e.g., some data), is designed as the least biased inference which accounts for known information, making it optimal in the sense of being neither overfitted nor incomplete for any given scenario. As a modelling principle, the constrained maximum entropy principle (CMEP) is canonically the soundest \cite{jaynes2, dill2}. For a complete review of the principles behind the CMEP, see \cite{giffon}. For convenience, we refer to this host of results as `Jaynes optimality.' A model is Jaynes optimal if it is optimal for a set of known unknowns, i.e., it produces the model which neither overfits those constraints nor ignores them, by assigning probability in accord to the preference induced by constraints \cite{Sakthivadivel2022b}. Contrast this with Bayes optimality, which is a model that provides a lower bound on classification error: Bayesian updates can be derived from maximum entropy (see \cite{jaynes-3} and references therein, or more recently, \cite{caticha}), so we assume these are closely related. Under the assumption that particles are Bayes optimal (perhaps justified by the complete class theorem that Bayes optimal agents do exist), they are Jaynes optimal, and \emph{vice-versa}. In the simpler case of conditional independence of inputs, for instance, it is known that na\"ive Bayes classifiers are CMEP models.

It stands to reason, then, that systems which model their environments use the principle of maximum entropy. Recall the use case driving the FEP: we need to use a principled modelling framework to form (scientific) models of things using principled modelling frameworks to form (recognition) models. In this sense, it would be ideal if the FEP is an elegant shorthand for applying maximum entropy to model things using maximum entropy. Recent literature \cite[Theorem 4.1]{Sakthivadivel2022b} has shown that this is precisely the case. 

Let us assume that a particle models its environment, justified by the deflationary approach to estimation-as-modelling discussed above. The FEP is what allows us to model this model; we want to prove that it is equivalent to the sort of model that such a particle actually could be expected to use. Take \eqref{vfe-eq} and negate it once, then maximise it instead of minimising it. Grouping the last two terms in the expression, we then have 
\[
-\int q(\eta; \sigma(\hat\mu_b), \Sigma_{\eta \mid b}) \log q(\eta; \sigma(\hat\mu_b), \Sigma_{\eta \mid b}) \dd{\eta} - \left(-\int q(\eta; \sigma(\hat\mu_b), \Sigma_{\eta \mid b}) \log p(\eta \mid b, \mu) \dd{\eta} -\log p(\mu, b) \right).
\]
Contrast this with maximising the entropy of the recognition model $q$ subject to the constraint that the surprisal of the environment given the particle is (on average) no greater than the intrinsic surprisal of the particular states: $- \E_q[\log q] - \left( \E_q[J(\eta)] - C \right)$ where $J(\eta) = -\log p(\eta \mid b, \mu)$ and $C = -\log p(\mu, b)$, 
and this proves the claim. That is to say, since maximising a positive quantity is the same as minimising a negative quantity, if things model their environments by using the CMEP, then the FEP is an equivalent (scientific) model of that (scientific) model. As such, the FEP is not merely a model of models, but is \emph{the} model of the models that all such systems that model their environments use, due to its relationship to the CMEP. The Jaynes optimality of the chosen recognition density in the FEP rests precisely on using constrained maximum entropy to choose that density.

The point of this section has ultimately been that, if particles use the CMEP to model their environments (which is reasonable given they are optimal in the sense discussed above) then understanding what our FEP-theoretic model does is like inhabiting the system and understanding what the particle does. Recall that by constructing a model of the particle modelling its environment, there exists an induced model of the states of the particle $\mu$ doing that modelling under $\sigma$. Next, we will prove that the induced model on internal states is also optimal, for largely the same reason: ourselves, as observers in an environment, also ought to employ the CMEP to model particles in \emph{our} environment\textemdash a symmetry which is brought to life by the FEP.

\section{Of maps of maps}

\subsection{The duality of FEP and CMEP}\label{dual-section}

There are two ways to model a particle modelling its environment: (i) to model the environment as though we are the particle modelling its environment (from its own first-person perspective), or, (ii) to model the particle holding a model of the environment. These perspectives reduce to `forming a model of external states as if we were the particle'\textemdash what we have called a map of the piece of the territory which behaves like a map, which allows us to understand the inference performed by other particles, which is in turn the core of the FEP\textemdash and `forming a model of the internal states which model external states,' the actual application of the FEP, which allows us to write down those inferences on paper. Paradoxically, these are equivalent statements, in that both `access' the generative model enailed by the system. However, as methods they go about that goal in radically different\textemdash even directly opposing\textemdash ways. Opposites that are equal in some sense are known in mathematics as adjoint pairs. 

It has long been remarked that the FEP-theoretic problems are equivalent to maximum entropy problems with a constraint, where the constraint is a log-probability function supplied by a generative model. Recent work has demonstrated that the FEP is, in a specific sense, mathematically equivalent to the CMEP. They are two sides of the same metaphorical coin. More precisely, they are dual to each other \cite{Sakthivadivel2022b, ramstead2022bayesian}; see in particular \cite[Theorem 4.2]{Sakthivadivel2022b}. The existence of a duality relationship between two structures means (informally) that those two structures are dual to each other, i.e., that they form an adjoint pair. Adjoint pairs are structures with identical `objects,' but where things that act on those objects act in an opposite manner to each other. Here, the adjunction is semi-obvious: the objects (internal and external states) are the same, but our model either (i) looks `outwards', from the particle's first-person perspective, by forming a model of beliefs and inferences about external states, or (ii) looks `inwards' at the system from a scientific (second person) perspective by forming a model of the dynamics of internal states in relation to external states. 

We can understand this as the following: suppose we have the recognition density $q(\eta \mid \mu)$, a parameterised family of proposed recognition models held by the particle, each of which is conditioned on the particular value of the internal state holding that belief. Then one value of $\mu$ yields the optimal model, $\mu = \hat\mu_b = \sigma^{-1}(\hat\eta_b)$. Our model of a thing which models its environment under maximum entropy can thus be formed by maximising the entropy of our own scientific model under the constraint that our model should model a system modelling its environment. This is the meta-Bayesian application of the constrained maximum entropy principle, where the constraints are supplied by ideal Bayesian observer assumptions (i.e., that the behaviour of the observed observer complies with the FEP \cite{daunizeau}). This simply states that our model of particular states should consider that the particle as an observer or estimator, and reproduce our intended aim of modelling the particle given our knowledge that the particle's behaviour is contingent on its model. Thus, we can maximise the following entropy functional:
\begin{equation}\label{dual-eq}
-\E_{p(\mu\mid b)}[\log p(\mu \mid b)] - \lambda \left( \E_{p(\mu\mid b)} J(\mu) - \Sigma_{\eta\mid b}\right)
\end{equation}
where $J(\mu) = [\mu - k]^2$ and $k = \sigma^{-1}(\hat\eta_b)$\textemdash in other words, a model of a system that (on average, though how often precisely is controlled by the Lagrange multiplier $\lambda$) holds an optimal model of its environment. Note that this particle also matches the variance of the environment, not just the mode, in order to model the full suite of environmental states. Notice also that we have switched the locations of $\eta$'s and $\mu$'s compared to \eqref{vfe-eq}, now maximising the entropy of the model of internal states modelling external states, rather than the model of external states. This reflects the duality referenced. Together, we implicitly have a maximum entropy model of the system holding a maximum entropy or FEP-theoretic belief.\footnote{For completeness, note that we often assume the synchronisation map applies to the conditional \emph{maximum a posteriori} estimate, such that the system infers the parameter controlling the placement of the peak of the probability density: the most likely state of the environment. Our maximum entropy model of this model assumes the particle is performing a Laplace approximation \cite{DaCosta2021}. Importantly, note that the log-probability constraint in \eqref{vfe-eq} is precisely the variance term in \eqref{dual-eq} under a Laplace approximation, meaning that \eqref{vfe-eq} and \eqref{dual-eq} are identical up to dualisation in that case. More general forms for \eqref{dual-eq} are possible which generalise beyond the Laplace assumption.}

This leads us to another, absolutely core, point\textemdash one which is central to the FEP itself, and hence to any candidate philosophy of the FEP and Bayesian mechanics. The duality we have leveraged arises from the fact that the boundaries of a particle are symmetrical. In turn, this means the environment of a particle is also an agent that models the particle. More precisely, the particle and environment each model the actions of the other on their mutually shared blanket states. Nature is constantly dissipating organised structures in the world, and the inference that is performed by the natural world is how best to dissipate the energy held in any such structure. The only way to avoid this is to mirror nature herself; this is the point of us modelling `things' as being unsurprising. This is the game that is characteristic of self-organisation, and only by understanding both sides can we understand self-organisation; such an attitude has been referred to as a relational approach to biological physics \cite{rosen}.

In the end, map and territory are simply dual perspectives; the fact that the FEP applies to us, and that we articulate models from our perspective as external observers, may often blind us to this duality. However, it underwrites the whole argument on offer: the FEP packages together an optimal model of a thing and an optimal model of the thing's model to achieve a new view on self-organisation. This is also what we mean by the FEP being the application of maximum entropy to things that maximise entropy, again making it unique amongst modelling methods. By modelling a particle modelling its environment, we achieve a dual model of the particle automatically, induced by the symmetry between external and internal states. However, a further uniqueness of the FEP is that, by emphasising the viewpoint of internal states, it is uniquely suited modelling \textit{action} and \textit{cybernetic systems}. The dual, CMEP-theoretic viewpoint is a good (i.e., Jaynes-optimal) account of the dynamics of particles that stay together, in virtue of looking at an anomalous `thing' that organises (i.e., the viewpoint of some scientist or heat bath), but does not include an account of action or policy explicitly \cite{Sakthivadivel2022b}. As indicated, this agent-centred perspective is what the FEP brings to the table.

That the FEP and the CMEP turn out to be two complementary perspectives on the physical existence of systems with particular partitions has profound implications, which have not been remarked upon previously in great detail (although see \cite{ramstead2022bayesian} for an exposition of the mathematical nature of this duality). FEP-theoretic models of complex physical systems, as we have seen, are mechanical theories, which provide us with an explanation (i.e., laws of motion) of the observable behaviour (or dynamics) of a system, from the point of view of a particle that self-organises (i.e., from the point of view of a `self'). Dually, the CMEP provides a description of self-organising systems, this time from the point of view of an observer, located in the external environment. Now, it is a mathematical fact that one can appeal to this duality to convert any FEP-theoretic model into an equivalent CMEP-theoretic one. More precisely, an FEP-theoretic model of belief updating (i.e., a model of the dynamics of beliefs, encoded by the internal states of a particle) can always be converted to a dual CMEP-theoretic model of systemic dynamics (i.e., a model of the dynamics of the states or paths of the system \emph{per se}, as opposed to the beliefs that they encode). Moreover if we have one, we implicitly have the other. This argument extends the dual aspect information geometry of \emph{internal states}\textemdash describing their thermodynamics \cite{crooks}\textemdash and the information geometry that arises when internal states are read as the sufficient statistics of Bayesian beliefs \emph{about external states} \cite{wiese}. 

Besides being a well-appreciated and established theory in physics, as we will go on to discuss, the relationship to the CMEP justifies the FEP as a statistical-mechanically optimal model of some unknown data-generating process. 

\subsection{FEP-theoretic models of encoded beliefs are CMEP-theoretically optimal models of systemic dynamics}\label{optimal-section-ii}

The effectiveness of the CMEP in statistical physics and beyond is not to be taken lightly, and the implications of this duality are far reaching. Some commentators have argued that the FEP is not special \cite{raja2021}: on this view, the FEP is merely one formal approach to the dynamics of complex systems, without any special epistemic status. The duality of FEP and CMEP negates this claim. Indeed, as we have outlined, maximum entropy inference is a way of arriving at the best model possible, mathematically, given our state of knowledge about the system being modelled. So, the duality of FEP and CMEP has a dramatic and deeply significant implication: given the duality of FEP and CMEP, we can state in full generality that a FEP-theoretic model of a particle’s belief updating is the optimal model of systemic dynamics, from the point of view of an external observer. We know that an FEP-theoretic model of a particle’s beliefs about a system simply is a maximum entropy model of the states of the system, albeit seen from another perspective. We know that maximum entropy inference enables us to arrive at the optimal model of that system, given our state of knowledge. The implication is that, given some constraints, the FEP-theoretic model is the optimal model of the physics of a system. This makes intuitive sense: due to the way that the generative model or Lagrangian is defined (i.e., over all the subsets of the systems), the FEP-theoretic model contains every ingredient needed to fully account for the manner in which subsets coupled to each other. 

Of course, practically, it does not suffice to write down a path of least surprisal. Minimally, without some additional specification, such a scientific model cannot be complete; since the boundary conditions (e.g., initial conditions and constraints) have to be added. More directly relevant to our present concerns, one must specify the form of the generative model or Lagrangian, which is often far from a trivial exercise. It is always a model in context. Indeed, it is the model or energy function which does most of the heavy lifting when it comes to modelling specific, empirical systems. We still need to find the appropriate set of constraints; what the technology of FEP/CMEP enables is doing the best inference possible, given our state of knowledge. Indeed, given that they are principles (i.e., mathematical structures with no empirical content), one cannot falsify the FEP/CMEP directly \cite{Andrews2021, ramstead2022bayesian}. This would be like attempting to falsify calculus or group theory: this constitutes a category error. One can only falsify specific models that have been derived from FEP/CMEP-theoretic technologies. When we arrive at incorrect results using these methods, the mistake usually lies in the application: the inference was optimal, given the non-optimal data set that we were using to constrain our inference procedure. What counts, in some sense, then, is the function that we put the model to use. Assuming we have adequate knowledge of the target system, and have written that knowledge down adequately as a set of sensible constraints, then the FEP-theoretic model is the optimal model; and if we have not done so, then we ought to be prepared for our model to not be optimal. The same problem has been known in Bayesian statistics for some time, and hence ought not to be particularly striking \cite{wolpert}.

These considerations serve to decisively defuse the second horn of the map-territory fallacy, which is that map and territory are always separate. Indeed, the FEP is not a mere model of physical processes: it is the optimal model given our state of knowledge, and therefore, poses absolute constraints on what it means to be a model of some physical process.

\section{Towards a philosophy of the FEP and Bayesian mechanics}

We now briefly summarise and elaborate on the points discussed above. We propose these few points as the foundations of a philosophy of Bayesian mechanics, which has yet to be constructed. 

\subsection{A map of map-like territories, and a map of possible maps: Towards a dynamical, nested representationalism}

In summary, we have argued that the map-territory fallacy, as it has been leveraged against the FEP (e.g., by \cite{Bruineberg2020, VanEs2020}) simply does not apply to FEP-theoretic modelling: it constitutes a fallacy, which we have called the \textit{map-territory fallacy fallacy}. Mathematically, there is no ambiguity or conflation of map and territory in FEP-theoretic modelling. Although the language used to narrate the mathematics of the FEP are at times ambiguous (which may have led to the confusion of some segments of the philosophical commentary and reporting on the FEP), the mathematical structure itself is not. That the target system itself \textit{is} a generative model or Lagrangian ought to be read as shorthand, to mean that the model that we deploy as scientists is meant to capture all the dependencies that exist within the system. The particle itself deploys a \textit{recognition model}, which its internal states encode; their dynamics are given in \textit{our} generative model, which is meant to capture the real dependence relations of the target system \cite{Ramstead2019enactive}. Applications of the FEP thus commit to a kind of `nested representationalism,' in the sense that it is a scientific model (implemented as a generative model of states or Lagrangian of paths), the content of which is precisely a mechanistic theory for the representational capacities of certain particular kinds of systems. Here, and in subsequent publications, we argue that the correct interpretation of the FEP is a dynamical, nested representationalist one.

Moreover, as we have seen, the FEP is not merely one possible way to model physical systems. Instead, it is the optimal model of systemic dynamics, given our state of knowledge. Thereby, the FEP/CMEP duality imposes ultimate constraints on what it means to be a physical model of some target system. Thus, FEP/CMEP provides us with core mathematical constraints on the optimality of modelling strategies. This determination of constraints on what it means to physically realise a model from within the formal constraints of the mathematics of physical modelling itself is reminiscent of Kant’s attempt to arbitrate the limits of reason from within the system of reason itself \cite{kant1908critique}; or again, Wittgenstein's attempt to formulate the structure of any proposition in general, within the language of propositional calculus \cite{wittgenstein2013tractatus}: it is an attempt to provide constraints on any possible meaningful maps, which arise from what it means to be a map or model of a physical process.

What is unique about the FEP/CMEP approach is that it is ultimately a model that is apt to model the modeller. Indeed, the Markov blanket is a \textit{symmetric} boundary between particle and environment. We see the world through our own Markov blanket; and any possible observer constitutes the environment for the particle that we are modelling. From the point of view furnished by the FEP/CMEP duality, our scientific models are attempts to infer what reveals itself vicariously through our Markov blanket. What we, as modellers, call systems or particles ultimately reflect what we, as modellers, can observe, and identify as persistent entities that sketch themselves out on our Markov blanket. 

This position can be illustrated by revisiting the map–territory fallacy through the lens of the FEP. The foregoing argument suggests that one should not ask whether the FEP is subject to the map-territory fallacy; rather, one should ask whether the map-territory fallacy subject to the FEP. With a careful and formal definition of map and territory, the answer is yes, in the following sense. One can associate the `territory' with the generative model describing the joint probability over causes and consequences, from the perspective of some `thing.' The `map' may be best ascribed to the recognition model in the sense that a map denotes a mapping; here, a map from internal states to Bayesian beliefs about external states. Note that for a `map' to exist in this sense, there has to be a partition of states: in the absence of a Markov blanket\textemdash which distinguishes internal states from external states\textemdash there can be no map, and \textit{therefore, no fallacy}. With these definitions in place, the map–territory fallacy reduces to a failure to distinguish between recognition and generative models. This distinction is precluded by the FEP, whose core claim is that the map depends in a lawful way on the territory\textemdash and that this mapping has existential implications, i.e., that there exists a particular partition or something that holds a recognition model about external states \cite{Ramstead2019enactive}. 

So, can there be a map-territory fallacy in applications of the FEP to form scientific models? Again, the answer is yes, with some interesting examples. In applications of the FEP, we are building recognition models (i.e., maps) of generative models (i.e., territories) that best explain the behaviour of something. To reify one of these maps is to falsely infer that our scientific recognition model is the true generative model. Examples of this in statistics could include the fallacy of classical inference \cite{lindley}; in which the false inferences rest upon using the wrong kind of (point null) hypotheses. Perhaps a more interesting example of false inference is psychopathology; namely, the type I errors when we falsely infer something is there when it is not (e.g., hallucinations and delusions) or type II errors when we falsely infer something is not there when it is (e.g., dissociative symptoms and neglect syndromes). In fact, there is a large literature in computational psychiatry \cite{precisionpsych, benrimoh, smith, linson, sterzer, ainley} that could be read as one kind of map–territory fallacy (i.e., conflating one's false inferences with reality)\textemdash a very special kind of fallacy that inherits from applications of the FEP. Holding an optimal map for the wrong territory is the same sort of failure to be a proper Bayesian that we discussed in Section \ref{optimal-section-ii}\textemdash not a failure of Bayesian inference \emph{per se}, but a misapplication of such an inference.

\subsection{On the theoretical appropriation of the FEP in philosophy}

The core publications on the FEP and Bayesian mechanics have mostly been in mathematical biophysics \cite{friston2012, Sakthivadivel2022b, Friston2019} and computational or theoretical neuroscience \cite{Friston2010, Friston2017graphical}. Since the middle of the last decade, he FEP has also generated much interest in philosophy \cite{clark2015surfing, Hohwy2014Predictive}. Despite these efforts, and with a few exceptions (e.g, \cite{Andrews2021, wiese}), there has been very little work on the philosophical commitments of the FEP per se; and, given that it has only emerged self-consciously over the last few years, there is only one publication to date discussing the philosophy of Bayesian mechanics \cite{ramstead2022bayesian}. Indeed, arguably, most extant philosophical work on the FEP starts from a given set of assumptions\textemdash for example, about the extended mind \cite{Kirchhoff2019}, the enactive approach to cognition \cite{Ramstead2019enactive}, or neo-Kantian, indirect accounts of perception \cite{Hohwy2017}\textemdash and then proceeds to present a reading of the FEP that fits as well as possible with those assumptions. This seems especially common in scholarship that is independently committed to philosophical theses that are, in many crucial ways, contradictory to the core commitments of the FEP, such as staunch anti-representationalism and the rejection of computational and information theoretic tools to study self-organising systems \cite{VanEs2020compu, hipolito2022enactive}. 

We do not think that this is conducive to understanding the theory and commitments of the FEP and Bayesian mechanics themselves (although we have been guilty of this ourselves, in some cases). Indeed, we believe that the variegated theoretical appropriation of the FEP is one central reason why there \textit{seem to be} several different interpretations or versions of the FEP\textemdash with wildly conflicting theoretical assumptions and broader philosophical commitments. We urge researchers in the tradition to pursue their examinations of the FEP on its own merits, setting aside questions about its compatibility with specific, independent philosophical perspectives (or the lack thereof).  

\section{Conclusion}

This paper has taken some first steps towards the development of a semi-axiomatic meta-theory and philosophy of the free energy principle (FEP) and Bayesian mechanics, providing a justification of its use in the modelling of physical systems, and an explanation of its scope in terms of first principles. We argued that, in describing `things' or `particles' as estimators or (deflated) representations of the systems to which they are coupled, the FEP-theoretic apparatus does not commit the map territory fallacy, i.e., it is false that the FEP reifies aspects of the metaphorical map (i.e., our scientific model), mistakenly taking them to be part of the territory (i.e., the target system that we wish to model). We have argued that this allegation itself constitutes a \textit{map-territory fallacy fallacy}. We have argued that, in distinguishing the generative and recognition models, FEP-theoretic modelling allows us to construct a map of that part of the territory that \textit{behaves as if it were a map}, without reification. Second, we have leveraged the duality of the FEP and the CMEP to argue that, in a Kantian or Wittgensteinian manner, the FEP also provides us with ultimate constraints on what it means to be an optimal model of physical processes. The FEP us thus, metaphorically, a map of any possible map whatsoever of, or held by, a physical system. 

We ought to celebrate the territory map mapping. The FEP is really, at its core, a principled approach to the formalisation of this mapping. The FEP is utterly unique as a `Jaynes-optimal' model of physical systems that model (i.e., are estimators of the statistics of) their environments. This restores convention to the FEP, via its relation to the CMEP: the FEP is not a strange theoretical beast residing in a faraway corner of theoretical biology, but rather, turns out to be a core principle of physical science. Our meta-theoretical approach to the FEP clarifies its role and scope as the uniquely ideal or optimal modelling tool for generic systems in statistical physics.

\subsection*{Acknowledgements}

We are thankful to Mahault Albarracin, Mel Andrews, Lancelot Da Costa, Chris Fields, James Glazebrook, Alex Kiefer, Gabriel Ren\'e, and Adam Safron, as well as the regular attendees of the Theoretical Neurobiology Group Meetings at University College London’s Wellcome Centre for Human Neuroimaging, for fruitful discussions that helped to shape the content of this paper.

\bibliographystyle{unsrt}
\bibliography{main}

\end{document}